\documentclass[twocolumn]{aastex7}
\usepackage{CJK}
\usepackage{lipsum}
\usepackage{amsmath}
\usepackage{gensymb}
\usepackage[dvipsnames]{xcolor}
\usepackage{natbib}
\usepackage{threeparttable}
\usepackage{hyperref}

\defcitealias{gaia2024}{Gaia24}

\newcommand{\msun}{M_\odot}

\newcommand{\ain}{a_{\text{in}}}
\newcommand{\aout}{a_\text{out}}

\newcommand{\ein}{e_{\text{in}}}
\newcommand{\eout}{e_\text{out}}

\newcommand{\nout}{n_\text{out}}

\renewcommand{\in}{\text{in}}
\newcommand{\out}{\text{out}}
\newcommand{\imut}{i_\text{mut}}

\newcommand{\revise}{\color{black}}

\begin{document}
\begin{CJK*}{UTF8}{gbsn}

\title{Exciting Stellar Eccentricity in Gaia BH3 via a Hidden Black Hole Binary}
\shorttitle{Excitation of Stellar Eccentricity in Gaia BH3}
\shortauthors{Hu et al.}

\author[0009-0007-9015-9451]{Qingru Hu (胡清茹)}
\affiliation{Department of Astronomy, Tsinghua University, 100084 Beijing, China}
\email{huqr24@mails.tsinghua.edu.cn}

\author[0000-0002-0643-8295]{Bin Liu (刘彬)}
\affiliation{Institute for Astronomy, School of Physics, Zhejiang University, 310027 Hangzhou, China}
\affiliation{Center for Cosmology and Computational Astrophysics, Institute for Advanced Study in Physics, Zhejiang University, 310027 Hangzhou, China}
\email{liubin23@zju.edu.cn}

\author[0000-0003-4027-4711]{Wei Zhu (祝伟)}
\affiliation{Department of Astronomy, Tsinghua University, 100084 Beijing, China}
\email{weizhu@tsinghua.edu.cn}

\correspondingauthor{Qingru Hu}
\email{huqr24@mails.tsinghua.edu.cn}
\correspondingauthor{Bin Liu}
\email{liubin23@zju.edu.cn}
\correspondingauthor{Wei Zhu}
\email{weizhu@tsinghua.edu.cn}

\begin{abstract}
We propose that the high eccentricity of the stellar orbit in Gaia BH3 system could be excited through a secular resonance effect if the inner dark object is, in effect, a tight and eccentric black hole binary (BHB). During the orbital decay of the inner BHB, the apsidal precession rate of the inner binary matches that of the outer stellar orbit, and this resonance advection can drive the outer eccentricity into some extreme values. 
For a Gaia BH3-like system, we show that a near equal-mass ($q=0.8$)
BHB with an initial semi-major axis of 1--3 au and an initial eccentricity $\gtrsim 0.95$ is able to excite the outer orbit to the observed value, leaving a current BHB with semi-major axis 0.25--0.5 au and eccentricity $\sim0.8$. 
The eccentric inner BHB imprints two observable signatures on the outer star: (1) short-term RV modulations with an amplitude $\lesssim 100$ m/s and (2) long-term apsidal precession with a rate $\lesssim 0.1\degree$/yr. Although neither of these is detected in the currently available astrometry and RV data, we show that these signals are detectable with the full Gaia astrometry data and dedicated high-precision and/or long-term RV observations.
Our work provides a new perspective on the dynamical formation of Gaia BH3, and the methodology is readily applicable to similar systems such as HD 130298, Gaia BH1, and Gaia BH2.

\end{abstract}



\section{Introduction}

{\revise To date, the \textit{Gaia} mission \citep{gaia2016} has discovered three dormant black holes residing in star-BH binaries: Gaia BH1 \citep{Chakrabarti2023AJ,gaiabh1_2023}, Gaia BH2 \citep{Tanikawa2023ApJ,gaiabh2_2023} and Gaia BH3 (\citealt{gaia2024}, hereafter Gaia2024). While the first two systems have $\sim 1 \msun$ stars around $\sim 10\msun$ BHs on moderately eccentric ($e\simeq 0.5$) and wide ($a\simeq 1-5$ au) orbits, the BH in Gaia BH3 has a high mass of $\sim33~\msun$ and is on a wider ($\simeq 16$ au) and highly eccentric ($e\simeq 0.73$) orbit around a metal-poor giant star of mass $M_\star\simeq0.76\,M_\odot$.}

{\revise The formation pathways of galactic BHs like Gaia BHs are highly uncertain and remain heatedly debated. The first possibility is that Gaia BHs are formed from isolated binary evolution. For Gaia BH1 and Gaia BH2, their orbits seem too wide to have formed through common envelope evolution in isolated binaries \citep{gaiabh1_2023,gaiabh2_2023}. For Gaia BH3, isolated binary evolution is able to produce such systems under certain conditions due to the very metal-poor companion star \citep{ElBadry2024OJAp, Iorio2024}. The second possibility is that Gaia BHs are assembled dynamically in dense star clusters \citep[e.g.][]{2018ApJ...855L..15K}. Previous simulations have demonstrated that dynamical assembly in an open cluster could produce Gaia BH1 and BH2 alike systems \citep[e.g.][]{2024MNRAS.527.4031T,2025MNRAS.538..243F}. The association of Gaia BH3 with {\revise the ED-2 stream \citep{Balbinot2024ED2stream}} may also point to a dynamical origin that the system could be assembled from a massive BH and a visible star in the progenitor globular cluster of the ED-2 stream \citep{daniel2024}.
}


{\revise The third possibility is that Gaia BHs are formed as hierarchical triples, with the visible star as the tertiary to an inner binary containing two massive stars that evolve into BHs eventually \citep[e.g.][]{gaiabh1_2023,Naoz2025ApJ}.} The measured high mass of Gaia BH3 makes it more likely that the dark object in Gaia BH3 could be composed of a black-hole binary (BHB)\footnote{We clarify that ``BHB'' in this work refers specifically to a binary system composed of two BHs, distinct from some literature where it denotes a star-BH binary.} rather than a single BH. 
The formation of BHB triple with a visible star is challenging due to the extreme expansion of BHB progenitors (typically $10^2-10^3\,R_\odot$), which generally disrupts triple-system stability. However, some low-metallicity binaries may avoid expansion \citep{Marchant2016}, enabling compact triples \citep{Vigna2021ApJ}. Alternatively, such systems can be effectively formed in low-metallicity open clusters \citep{Tanikawa2025OJAp}.

If Gaia BH3 indeed hosts a BHB, non-trivial dynamical effects emerge and could significantly influence the evolutionary history of the system \citep[e.g.,][]{Liu2015MNRAS,Naoz2017AJ....154...18N,Antonini2017ApJ...841...77A}. Recent studies demonstrated that in coplanar triples containing a merging BHB, the outer star may experience extreme eccentricity growth when its apsidal precession rate matches that of the inner BHB \citep[e.g.,][]{Liu2022PhRvD,liu2024extreme}. 
We show in this work that such an ``apsidal precession resonance'' capture could naturally explain both the high eccentricity of the stellar orbit and the large mass ($\sim 33~\msun$) of the dark object in the Gaia BH3 system.

The BHB hypothesis can be tested via high-precision astrometric and/or RV observations of the visible star. An inner BHB would imprint two distinct signatures on the visible star: (1) short-term RV modulations induced by the orbital motion of the binary \citep{morias2008wobble,hayashi2020strategy}, and (2) long-term orbital precession from secular perturbations \citep{Liu2015PhRvD,hayashi2020radial}. The recent constraints on the binarity of Gaia BH1's dark companion by \citet{nagarajan2024espresso} demonstrate the feasibility of detecting such inner BHBs hidden in Gaia BH systems.

In this Letter, we first present in Section~\ref{sect:2} the possibility that the high eccentricity $e\simeq0.73$ of the stellar orbit in Gaia BH3 system could have been excited by an inner merging BHB through apsidal precession resonance capture, and then assess in Section~\ref{sec:detectability} the detectability of such BHBs with high-precision astrometric and RV observations. A brief discussion is given in Section~\ref{sec:discussion} on the influence of non-coplanar orbits and the applicability to other detached eccentric BH systems. Finally, we summarize our findings in Section~\ref{sec:conclusion}.

\section{Apsidal Precession Resonance}\label{sect:2}

If the $33 \, \msun$ companion in Gaia BH3 system is actually a black-hole binary (BHB) instead of a single BH, we will show in this section that, under certain initial configurations, the inner BHB can excite the eccentricity of the outer stellar orbit from near circular to the observed value ($\gtrsim 0.73$) {\revise within the system age ($\sim13$ Gyr) through the apsidal precession resonance. In this secular resonance mechanism, the 1st-order post-Newtonian potential and the gravitational wave (GW) radiation for the inner BHB, and the Newtonian perturbation up to octupole level between the inner BHB and the outer star are taken into account \citep[e.g.,][]{Liu2022PhRvD,liu2024extreme}. In this study, we do not include the hexadecapole level perturbation because it produces an apsidal precession on a much longer timescale \citep{PhysRevD.96.023017}, and leave its modification on the apsidal precession resonance for a future investigation.}

We assume here a coplanar hierarchical triple system in order to better understand the dynamical evolution and leave the more complicated situation of inclined systems to Section~\ref{subsect:noncoplanar}.
The inner binary consists of masses of $m_1$ and $m_2$ ($m_1 \geq m_2$), with a total mass of $m_{12}\equiv m_1+m_2 = 32.70~\msun$ from \citetalias{gaia2024}. The mass ratio of the inner binary is defined as $q\equiv m_2/m_1$ ($0<q\leq1$). The mass of the tertiary star is fixed at $m_\star=0.76~\msun$ (\citetalias{gaia2024}). The semi-major axes and eccentricities of the inner and outer binaries are denoted by $\ain$, $\aout$, and $\ein$, $\eout$, respectively. The semimajor axis of the outer binary takes the observed value $a_\out=16.17$ au \citepalias{gaia2024} throughout the paper, as $\aout$ is conserved under the double-averaged formalism in the secular regime. 
Since we focus on coplanar triples, we use the longitudes of periapsis,\footnote{{\revise The longitude of periapsis is generally defined as $\varpi=\omega+\Omega$, and could be used to specify the direction of periapsis for low-inclination orbits.}} $\varpi_\in$ and $\varpi_\out$, to specify the apsidal lines of the inner and outer binaries, respectively.

\begin{figure}[t]
    \centering
    \includegraphics[width=\linewidth]{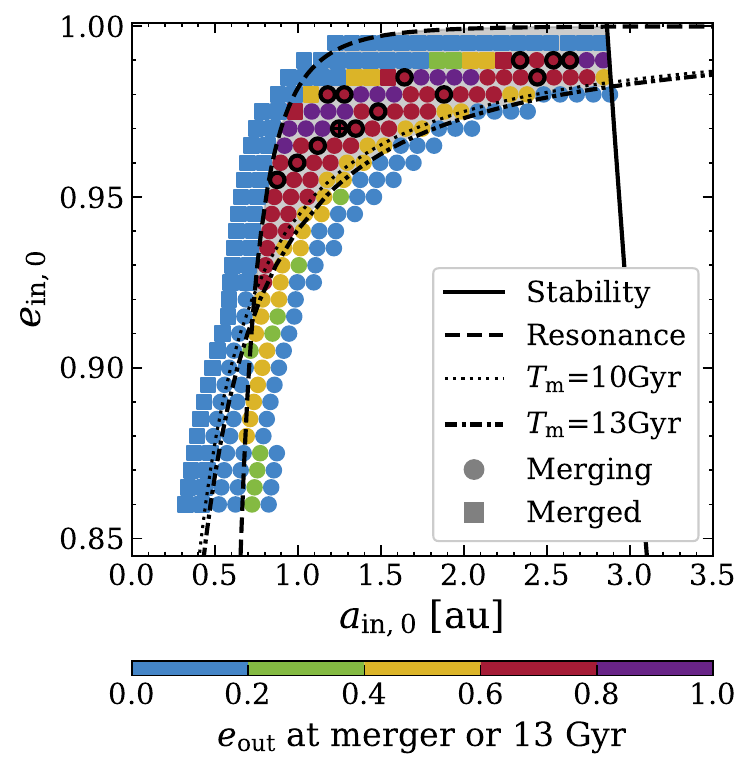}
    \caption{Valid initial configurations and the eccentricity of the outer binary at 13~Gyr. The enclosed region by the three conditions---stability (solid curve), resonance (dashed curve), and merger time (dotted and dashed-dotted curves, for two chosen ages)---indicates the initial BHB configurations that could potentially undergo apsidal precession resonance. Colored circles and squares represent the initial configurations used in our numerical integrations, with face colors corresponding to the eccentricity of the outer binary at 13~Gyr. Circles denote systems that have not merged after 13~Gyr, whereas squares represent those that have merged. Initial parameters of inner BHBs consistent with the observations after 13~Gyr of evolution are highlighted with black open circles. The black plus symbol marks the example system whose time evolution is shown in Figure~\ref{fig:time-evolution}.}
    \label{fig:1}
\end{figure}

We first identify the initial configurations that could potentially undergo the apsidal precession resonance by taking into account the resonance-capture condition, the orbital decay timescale due to gravitational radiation, and the stability of the coplanar triple. Note that all parameters in the initial state are denoted with an extra subscript ``0". Below, we choose a near equal-mass ($q=0.8$) BHB for illustration purposes. Results corresponding to other values of mass ratio $q$ are given in Appendix~\ref{appendix:1}. The initial eccentricity of the outer binary is set to $e_\text{out,0}=10^{-4}$ as an example of a near circular orbit.

In order for the apsidal precession resonance capture to happen, the parameter
\begin{equation}\label{equ:resonance}
    \gamma \equiv \frac{\dot{\varpi}_\text{out}}{\dot{\varpi}_\text{in}}
\end{equation}
must cross unity from above to below during the orbital decay of the inner BHB \citep{liu2024extreme}, {\revise where $\dot{\varpi}_\in$ ($\dot{\varpi}_\out$) denotes the apsidal precession rate of the inner (outer) binary. The dominant component in $\dot{\varpi}_\in$ is from the 1st-order post-Newtonian effect}
\begin{equation}
    \dot{\varpi}_\text{in,GR} = \frac{3G^{3/2}m_{12}^{3/2}}{c^2\ain^{5/2}(1-\ein^2)},
\end{equation}
{\revise and $\dot{\varpi}_\text{out}$ is dominated by the quadrupole-level secular perturbation from the inner BHB }\citep{Liu2015PhRvD}
\begin{equation}\label{equ:precession}
    \dot{\varpi}_\text{out,quad} = \frac{3}{8} \left( \frac{\ain}{\aout} \right)^2 \nout \frac{m_1m_2}{(m_1+m_2)^2} \frac{2+3\ein^2}{(1-\eout^2)^2},
\end{equation}
where $\nout$ is the mean motion of the outer orbit.
{\revise The critical condition of $\dot{\varpi}_\text{out,quad}=\dot{\varpi}_\text{in,GR}$ is plotted as the dashed line in Figure~\ref{fig:1}.}

The second condition is given by the orbital decay of the inner binary due to gravitational radiation, which changes both $\dot{\varpi}_\in$ and $\dot{\varpi}_\out$. 
The merger time due to gravitational radiation of a binary with initial semi-major axis $a_{\in,0}$ and eccentricity $e_{\in,0}$ is given by (e.g., \citealt{Peters1964})
\begin{equation}
\begin{aligned}
    T_\text{m} &\simeq \frac{10^{13} \ \text{yr}}{f(e_{\in,0})} \left(\frac{a_{\in,0}}{0.5 \ \text{au}}\right)^{4} \left(\frac{m_{12}}{30 \ M_\odot}\right)^{-3} \frac{(1+q)^2/q}{4}, \\
    f(e_{\in,0})&\equiv \ \frac{1}{(1-e_{\in,0}^2)^{7/2}} \left( 1+\frac{73}{24}e_{\in,0}^2+\frac{37}{96}e_{\in,0}^4 \right).
\end{aligned}
\end{equation}
We show the merger time of $T_\text{m}=10$ Gyr for a Gaia BH3-like system with different initial configurations as the dotted curve in Figure~\ref{fig:1}. Given the uncertainty in the system age \citepalias[12--14 Gyr,][]{gaia2024}, we also plot the merger time of $T_\text{m}=13$ Gyr as the dotted-dashed curve in the same figure, which is not too different from the $T_\text{m}=10$ Gyr curve. BHBs with initial semi-major axis and eccentricity above these curves have experienced significant orbital decay within the age of the system and thus may get locked into the apsidal precession resonance.

The third condition is the stability of the coplanar triple system. We choose the dynamical instability condition for hierarchical triples given by \citet{mardling2001tidal}
\begin{equation}\label{equ:stable}
    \frac{\aout}{\ain} > 2.8 \left[ \left(1 + \frac{m_\star}{m_{12}} \right) \frac{1+\eout}{(1-\eout)^3} \right]^{2/5} \left( 1-0.3\frac{i_{\text{mut}}}{180 \degree} \right),
\end{equation}
where $i_\text{mut}$ is the mutual inclination between the inner and outer binaries and is set to zero for our coplanar triples. The threshold of dynamical instability for a Gaia BH3-like system is shown as the solid black curve in Figure~\ref{fig:1}.

The region bounded by the three conditions given above in Figure~\ref{fig:1} is the set of valid combinations of initial semi-major axis and eccentricity for the inner BHB that could potentially undergo apsidal precession resonance capture. 
For Gaia BH3-like systems, the mechanism requires an inner BHB with an initially highly eccentric orbit (see also \citealt{liu2024extreme}).

For BHBs with initial configurations within the identified parameter space, we then investigate their dynamical evolution in order to compare with the observed properties of the Gaia BH3 system. We select 118 sets of $(a_{\in,0}, e_{\in,0})$ within the allowed parameter space; for comparison purposes, we also select 135 combinations of $a_{\in,0}$ and $e_{\in,0}$ that fall beyond the previously identified parameter space. The chosen parameter combinations are shown as colored circles/squares in Figure~\ref{fig:1}. For each combination of $a_{\in,0}$ and $e_{\in,0}$, we integrate the double-averaged secular equations of motion \citep{Liu2015MNRAS} for up to 13 Gyr or merger of the inner BHB to obtain the final (i.e., present) orbital parameters of the system. Those systems that have experienced BHB merger within 13 Gyr are marked as squares in Figure~\ref{fig:1}, while those that have not are marked as circles. The final orbital eccentricity of the outer binary at 13 Gyr or merger of the BHB is indicated by the color scale in Figure~\ref{fig:1}. About one-third of the systems (38 out of 118) with initial conditions in the gray-shaded region get caught in the apsidal resonance, and the final $e_\out$ reaches higher than the observed value $e_\text{obs}\simeq0.73$. Taking into account the stability of the coplanar triple, only 14 out of the 38 systems remain compatible with observations, which are highlighted by black open circles in Figure~\ref{fig:1}.

\begin{figure}[t]
    \centering
    \includegraphics[width=\linewidth]{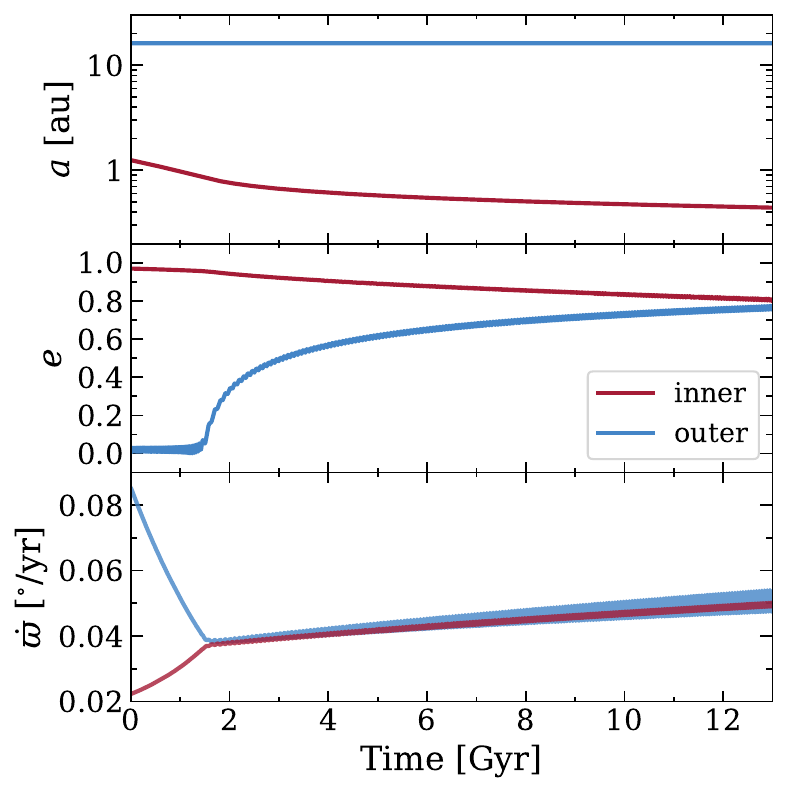}
    \caption{Time evolution of the semi-major axes $a$, eccentricities $e$ and apsidal precession rates $\dot{\varpi}$ of inner and outer binaries. This system is initialized with $a_{\in,0}=1.25$ au, $e_{\in,0}$=0.97, and $\varpi_\in=\varpi_\out=0\degree$, as highlighted by the black plus symbol in Figure~\ref{fig:1}. The red (inner) and blue (outer) curves represent the numerical solutions of the double-averaged secular equations as detailed in \citet{Liu2015MNRAS}.}
    \label{fig:time-evolution}
\end{figure}

Figure~\ref{fig:time-evolution} shows the time evolution of an example system with $a_{\in,0}=1.25$ au and $e_{\in,0}=0.97$ whose initial configuration is marked out with a black plus symbol in Figure~\ref{fig:1}. 
In the beginning, the apsidal precession rate of the outer binary is larger than that of the inner binary. As the inner BHB undergoes orbital decay due to gravitational radiation, the apsidal precession rates of the inner and the outer binaries converge, and the system is captured in the apsidal precession resonance at $\sim1.8$ Gyr. Since then, the eccentricity of the outer stellar orbit starts to increase due to the enhanced angular momentum exchange between the inner and the outer binaries. The system remains locked in the apsidal precession resonance until the end of the integration, and the eccentricity of the outer star increases to $e_\out \approx 0.76$ at 13 Gyr.

\begin{figure}[t]
    \centering
    \includegraphics[width=\linewidth]{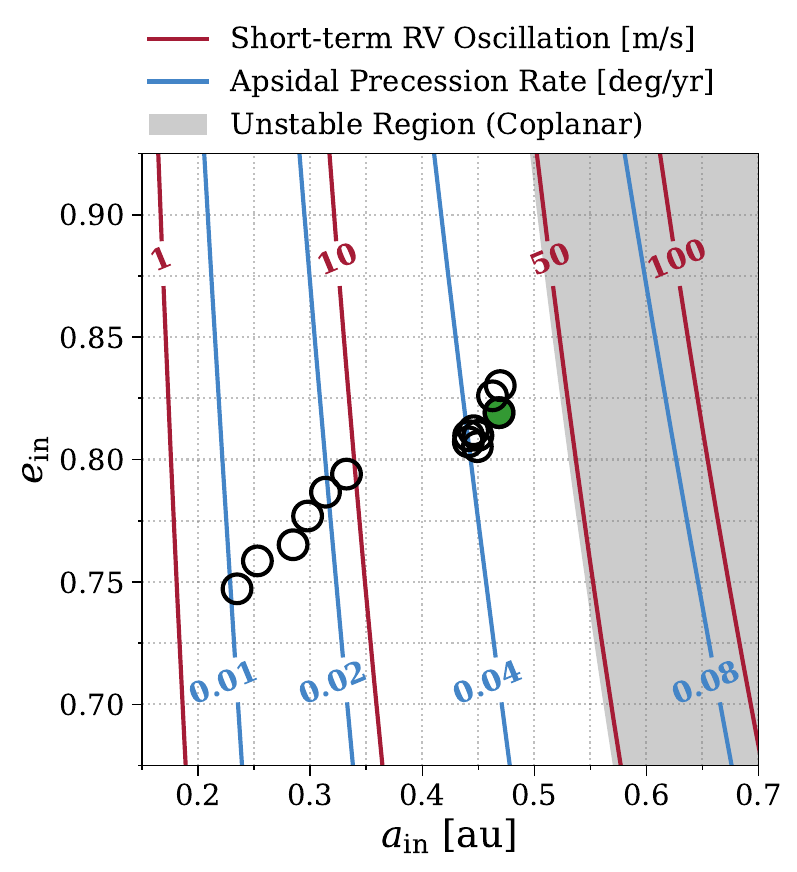}
    \caption{The black open circles indicate the present orbital parameters $\ain$ and $\ein$ of inner BHBs that are able to reproduce the observed eccentricity of the stellar orbit in Gaia BH3 and remain stable at the end of the integration. The green-filled circle highlights the final state of the example system, whose time evolution is shown in Figure~\ref{fig:time-evolution}. The red contours show the semi-amplitude of short-term RV oscillations (Equation~\eqref{equ:rv}) due to the orbital motion of the inner BHB, and the blue contours show the expected apsidal precession rate of the stellar orbit (Equation~\eqref{equ:precession}). The gray-shaded region indicates the unstable region of a coplanar triple with $\eout=0.73$ according to Equation~\eqref{equ:stable}.}
    \label{fig:signal}
\end{figure}

For those 14 configurations that successfully reproduce the observed $e_\out$ and remain stable at 13 Gyr, we show in Figure~\ref{fig:signal} the final eccentricity and semi-major axis of the inner BHB. 
These are the possible present states of the inner BHB if we assume that the eccentricity of the outer star is excited by apsidal precession resonance. The detectability of such BHB in the Gaia BH3 system is investigated in the next section.

\section{Detectability in Astrometry and RV} \label{sec:detectability}

\subsection{Short-term RV Modulations}

The semi-amplitude of short-term RV oscillations due to an inner binary inside a coplanar triple {\revise that takes into account the eccentricities of the inner and outer binary} can be estimated as \citep[e.g.,][]{morais2011AAA,hayashi2020strategy}
\begin{equation}\label{equ:rv}
    K_{\text{short}} \approx \frac{m_1 m_2}{m_{12}^2} \sqrt{\frac{m_{12}+m_\star}{m_{12}}} \left(\frac{\ain}{\aout}\right)^{7/2} g(e) K_\text{circ}.
\end{equation}
Here $K_\text{circ}$ is the characteristic RV semi-amplitude of the tertiary star around the center of mass of the inner binary 
\begin{equation}
    K_\text{circ} \equiv \frac{m_{12}}{m_{12}+m_\star} \aout \nout \sin i_\star ,
\end{equation}
and $g(e)\equiv [(1+\ein)/(1-\eout)]^{7/2}$ is an empirical factor that takes into account the eccentricities of the inner and outer orbits. 
The red lines in Figure~\ref{fig:signal} show the contours of relevant values of $K_\text{short}$ for the potential BHBs in the Gaia BH3 system. 
It is also worth noting that the actual RV deviation from a Keplerian motion also depends on the sampling cadence and strategy of the RV observations and could be 10--100 times larger than the analytic approximation given by Equation~\eqref{equ:rv}, as demonstrated by the simulations in \citet{Hayashi2023ApJ} for Gaia BH1 and Gaia BH2 systems.\footnote{{\revise Note that the definition of $K_\text{short}$ in \citet{Hayashi2023ApJ} does not include the eccentricity factor $g(e)$.}}
The maximum deviation appears around the periastron passage, and the next event for Gaia BH3 is in October of 2029 according to the orbital solution of \citetalias{gaia2024}.

The published RV measurements of Gaia BH3 are very sparse and not enough to exclude any of the potential BHB configurations in Figure~\ref{fig:signal}.

\subsection{Long-term Apsidal Precession}
The apsidal precession rate of the outer star induced by the secular perturbation from the inner binary is given by Equation~\eqref{equ:precession}. The blue curves in Figure~\ref{fig:signal} show the contours of a few apsidal precession rates relevant to the BHB parameters of Gaia BH3. The inner BHBs relevant to apsidal precession resonance mechanism are detectable if the apsidal precession rate can be measured at the level of $\sim0.01\ {\rm deg} \ {\rm yr}^{-1}$. {\revise We employ $\omega_\out$ to denote the pericenter of the outer star relative to the ascending node of its orbital plane and the sky plane. Although $\varpi_\out$ and $\omega_\out$ may differ in value, the time derivatives of both ($\dot{\varpi}_\out$ and $\dot{\omega}_\out$) equal the physical apsidal precession rate.}

In this subsection, we use the Gaia epoch astrometry and RV observations provided in \citetalias{gaia2024} to constrain $\dot{\omega}_\out$. The detectability of the apsidal precession by additional observations is also estimated. 

\begin{table*}[t]
\caption{Constraints on the apsidal precession (AP) model from the astrometry data alone, RV data alone, and the joint data set.}\label{tab:precess}
\begin{center}
\begin{tabular}{lccc}
\hline\hline
Parameter & Astrometry & RV & Astrometry \\
   & +AP & +AP & +RV+AP \\
\hline
$\Delta \alpha^{*}$ (mas)\tablenotemark{\tiny{a}} & $4.05 \pm 0.21$ & --- & $4.24 \pm 0.15$  \\
$\Delta \delta$ (mas)\tablenotemark{\tiny{a}} & $2.60 \pm 0.18$ & --- & $2.43 \pm 0.11$  \\
$\Pi$ (mas) & $1.680 \pm 0.009$ & --- & $1.686 \pm 0.007$ \\
$\mu_{\alpha^{*}}$ (mas/yr) & $-28.09 \pm 0.23$ & --- & $-28.30 \pm 0.15$  \\
$\mu_{\delta}$ (mas/yr) & $-155.5 \pm 0.3$ & --- & $-155.2 \pm 0.2$  \\
$P$ (years) & $11.6 \pm 0.3$ & 11.5 (fixed) & $11.48 \pm 0.23$  \\
$T_{\rm p}$ (JD-J2017.5) & $0.658 \pm 0.003$ & $0.664 \pm 0.025$ & $0.658 \pm 0.003$  \\
$a_\text{com}$ (au) & $16.50 \pm 0.41$ & 16.05 (fixed) & $16.12 \pm 0.20$  \\
$e_\out$ & $0.726 \pm 0.006$ & 0.726 (fixed) & $0.725 \pm 0.005$  \\
$i$ (deg) & $110.51 \pm 0.12$ & 110.66 (fixed) & $110.61 \pm 0.09$  \\
$\Omega$ (deg) & $136.29 \pm 0.15$ & --- & $136.19 \pm 0.11$ \\
$\omega_{\rm p}$ (deg) & $77.89 \pm 0.17$ & $77.7 \pm 1.0$ & $77.78 \pm 0.15$  \\
$\dot{\omega}_\out$ (deg/yr) & $-0.5 \pm 0.4$ & $-0.1 \pm 0.5$ & {\revise $-0.13 \pm 0.36$}  \\
$\gamma_{\rm spec}$ (km/s) & --- & $-357.1$ (fixed) & $-357.2 \pm 0.3$  \\
$\chi^2_\text{dof}$ & 0.92 & 1.12 & 0.98  \\
\hline
\end{tabular}
\end{center}
\tablenotetext{\tiny{a}}{The reference position for Gaia BH3 is $(\alpha_\text{ref},\delta_\text{ref})=(294\fdg82784900557243,~ 14\fdg930918410309376)$, while the reference time is J2017.5 (JD 2457936.875) according to \citetalias{gaia2024}.}
\end{table*}

The astrometric model with a constant apsidal precession rate can be constructed by replacing the original constant $\omega_\out$ with
\begin{equation}
    \omega_\out(t)= \omega_{\rm p} + \dot{\omega}_\out(t-T_{\rm p}),
\end{equation} 
where $T_{\rm p}$ and $\omega_{\rm p}$ are the time of periastron and the argument of periastron at $T_{\rm p}$, respectively. The modified astrometric model has 13 free parameters: small offsets in equatorial coordinates ($\Delta \alpha^\star, \Delta \delta$), proper motions ($\mu_{\alpha^\star}, \mu_\delta$), parallax ($\Pi$), orbital period ($P$), the time of periastron ($T_{\rm p}$), distance from the center of mass ($a_\text{com}$), eccentricity ($e_\out$), the orbital inclination ($i$), the longitude of ascending node ($\Omega$), the argument of periastron at $T_{\rm p}$ ($\omega_{\rm p}$), and a constant apsidal precession rate ($\dot{\omega}_\out$). We fit the Gaia epoch astrometric data from \citetalias{gaia2024} using our modified model, and the results are listed in Table~\ref{tab:precess}. Specifically, the apsidal precession rate is measured to be $-0.5\pm0.4\,{\rm deg\,yr}^{-1}$. 

The RV model with a constant apsidal precession rate can be derived from the definition of RV \citep[e.g.,][]{2019A&A...623A..45C}
\begin{equation}
\begin{aligned}
\text{RV} = K_\text{circ} &\left[ \cos(\nu(t)+ \omega_\out(t)) + e_\out \cos \omega_\out(t) \right. \\
    &\left. +\frac{\dot{\omega}_\out(1-e_\out^2)^{3/2}}{n_\out} \frac{\cos(\nu(t) + \omega_\out(t))}{1+e_\out\cos\nu(t)}  \right] \\
    {\revise + \gamma_{\rm spec}},
    \end{aligned}
\end{equation}
{\revise where $\nu$ is the true anomaly, and $\gamma_{\rm spec}$ is the spectrograph-dependent additive RV offset. The modified RV model has a total of 8 parameters ($P$, $T_{\rm p}$, $a_\text{com}$, $e_\out$, $i$, $\omega_{\rm p}$, $\dot{\omega}_\out$, $\gamma_{\rm spec}$), with the first 7 being identical to those in the modified astrometric model.} Because the available RV measurements are sparse, we choose to fix the parameters irrelevant to the periastron ($P$, $a_\text{com}$, $e_\out$, $i$, $\gamma_{\rm spec}$) to their best-fit values from a modeling without precession to the joint data of astrometry and RV, and then fit for the precession rate. The resulting constraint on the apsidal precession rate, also listed in Table~\ref{tab:precess}, deviates and yet is consistent with zero ($-0.1\pm0.5 \,{\rm deg\,yr}^{-1}$).

For completeness, we have also done a modeling of the joint dataset by simply adding the log likelihoods from the astrometry and RV data together. This combined solution is listed in the last column of Table~\ref{tab:precess}. The constraint on the apsidal precession rate is only improved by a small margin to $-0.13\pm0.36\,{\rm deg \, yr}^{-1}$.
The uncertainty remains significantly larger than the expected signal from an inner BHB ($<0.08\degree$/yr) as shown in Figure~\ref{fig:signal}, so a BHB inside the Gaia BH3 system cannot be ruled out by the currently available observations.

Below, we provide a simple estimate of how better constraints can be obtained with future observations. To the leading order, a constant apsidal precession model introduces a signal whose amplitude is $\sim \dot{\omega}_\out \Delta T$ smaller compared to the RV/astrometry signal of the Keplerian model. Here $\Delta T$ is the observational baseline. 
For astrometry, the uncertainty on the apsidal precession rate $\dot{\omega}_\out$ can be estimated as
\begin{equation}
    \sigma_{\dot{\omega}_\out} \simeq 0.03\ \text{deg\,yr}^{-1} \left(\frac{\sigma_\text{AL}}{0.14 \,\text{mas}} \frac{27 \, \text{mas}}{\rho} \frac{10 \, \text{yr}}{\Delta T}\right),
\end{equation}
Here $\sigma_\text{AL}$ is the uncertainty of the astrometric measurement in the along-scan (AL) direction, $\rho$ is the amplitude of the astrometry signal of the Keplerian motion. The evaluation is performed on the known properties of Gaia BH3 system ($\sigma_\text{AL}=0.14$ mas, $\rho=27$ mas, \citetalias{gaia2024}) at the end of the Gaia mission ($\Delta T=10$ yr).

Similarly, the uncertainty on the apsidal precession rate $\dot{\omega}_\out$ from long-term RV observations can be estimated as 
\begin{equation}
    \sigma_{\dot{\omega}_\out} \simeq 0.01\, \text{deg\,yr}^{-1} \left( \frac{\sigma_\text{RV}}{100 \, {\rm m\, s}^{-1}} \frac{50 \, {\rm km\, s}^{-1}}{K_{\rm circ}} \frac{10 \, \text{yr}}{\Delta T} \right),
\end{equation}
where $\sigma_\text{RV}$ is the precision of a single RV measurement. The above estimation is validated with simulated high-precision ($\sigma_\text{RV}=100$m/s, achievable with current instruments on Gaia BH3, \citetalias{gaia2024}) RV observations of five measurements per year for 10 years.

Therefore, both Gaia astrometry and long-term RV measurements are promising in constraining the apsidal precession rate induced by a BHB inside Gaia BH3 and testing the secular resonance hypothesis.

\section{Discussion} \label{sec:discussion}

\begin{figure}
    \centering
    \includegraphics[width=\linewidth]{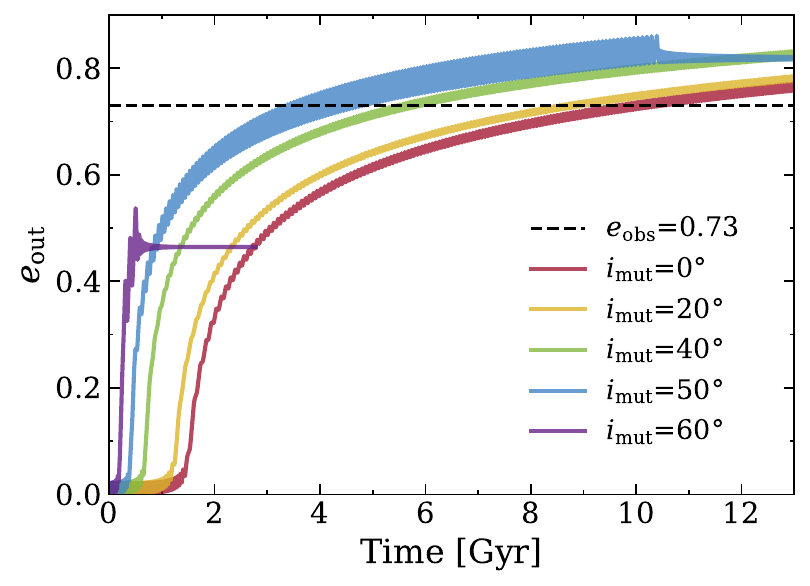}
    \caption{Time evolution of the outer eccentricity in misaligned triples with various mutual inclination angles ($\imut$). The systems are initialized with $a_{\in,0}=1.29$ au, $e_{\in,0}$=0.97, $\Omega_\in=\Omega_\out=0\degree$, $\omega_\in=0\degree$, and $\omega_\out=180\degree$, similar to that in Figure~\ref{fig:time-evolution}, except for the mutual inclination, which now varies from $0\degree$ to $60\degree$. The black dashed line marks the observed value of the stellar eccentricity ($e_\text{obs}\approx 0.73$).}
    \label{fig:inclined-triples}
\end{figure}

\begin{figure*}
    \centering
    \includegraphics[width=\linewidth]{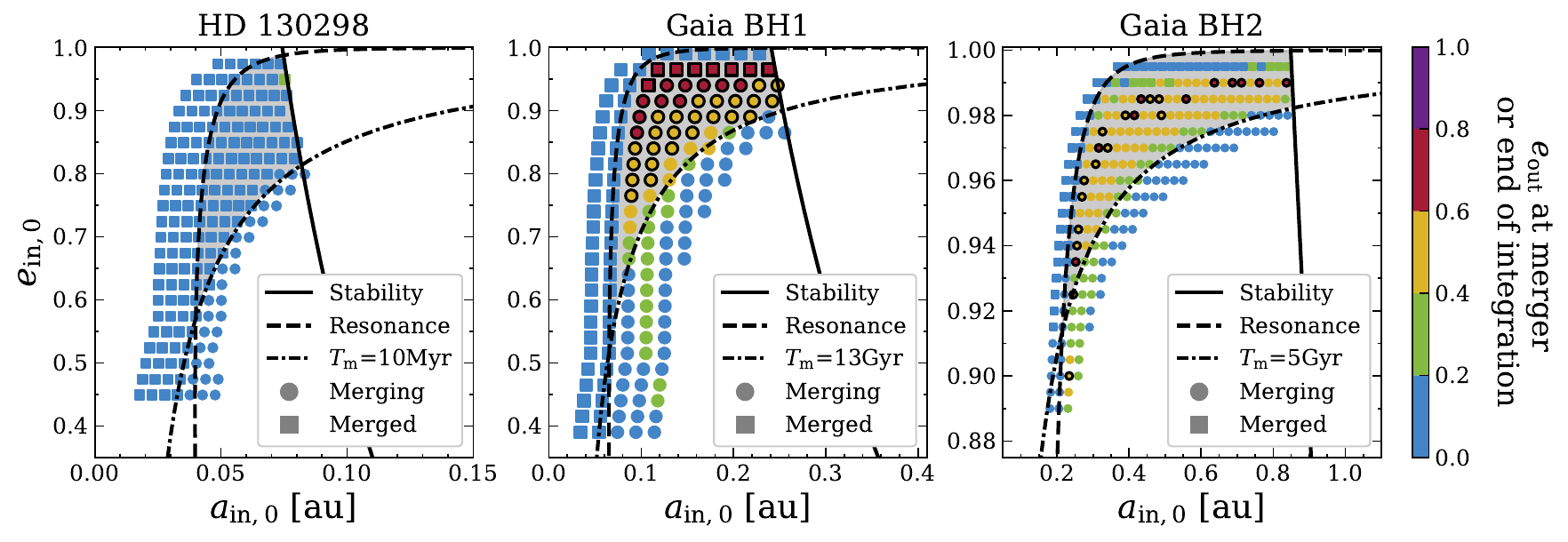}
    \caption{Similar to Figure~\ref{fig:1} but for three different BH systems (HD 130298, Gaia BH1, and Gaia BH2) {\revise based on relevant parameters as detailed in subsection~\ref{subsect:4.2}}.}
    \label{fig:6}
\end{figure*}

\subsection{Misaligned Triples}\label{subsect:noncoplanar}
Our investigation has thus far assumed the BHB inside Gaia BH3 to be coplanar with the outer star. We now discuss the implications of misaligned configurations.

For moderate mutual inclinations ($\imut \lesssim 50^\circ$) between the inner and the outer orbits, the second and the third conditions given in Section~\ref{sect:2} are largely unchanged. The condition for resonance capture is revised, and no simple analytical result can be derived, but our general conclusion is largely unaffected. 
One example is shown in Figure~\ref{fig:inclined-triples}, which compares the time evolution of the outer eccentricity for the same inner BHB as in Figure~\ref{fig:time-evolution} with various mutual inclinations ($\imut=0,20,40,50,60\degree$).

The eccentricity of the outer stellar orbit is securely excited by the inner BHB through the apsidal precession resonance mechanism, as long as the mutual inclination $\imut\lesssim50\degree$. 
The larger the mutual inclination, the earlier the system gets captured into apsidal precession resonance. This is because the critical $a_\in$ that determines the secular resonance shifts with $\imut$ \citep[more examples in][]{Liu2015PhRvD,Liu2020PhRvD.102b3020L,Liu2022PhRvD}. The final eccentricity of the outer stellar orbit also increases with increasing $\imut$. For large mutual inclinations ($\imut\gtrsim 50\degree$), the dynamics of the tertiary star is largely determined by the inverse von Zeipel-Lidov-Kozai \citep[ZLK,][]{von1910application,1962Lidov,1962AJKozai} oscillations, and the eccentricity excitation due to apsidal precession resonance is suppressed \citep{Naoz2017AJ....154...18N,Vinson2018MNRAS.474.4855V}. {\revise Besides, a highly-inclined tertiary star speeds up the merger of the inner BHB (e.g., the inner BHB merges at $\sim 3$ Gyr for the case of $\imut=60\degree$ in Figure~\ref{fig:inclined-triples}).}

\subsection{Other BHs in Detached Eccentric Binaries}\label{subsect:4.2}

We choose to study the Gaia BH3 system for its large BH mass and very eccentric orbit, the combination of which makes it reasonable to consider the dynamical origin of the stellar eccentricity in the BHB scenario. The studied mechanism is nevertheless generally applicable to other BH--star systems with detached and eccentric ($e\gtrsim0.4$) orbits. Below, we discuss briefly three other known systems that fall into such a category.\footnote{The recently identified Be star--black hole binary candidate ALS 8814 \citep{2025ALS8814} exhibits complex spectral variability and shows indications of a luminous companion \citep{2025OJApEl-Badry}. Due to these complexities, we do not provide a detailed discussion of the ALS 8814 system in this work.}

\begin{itemize}

\item \textbf{HD 130298} contains a dark object of mass $\gtrsim 7.7~\msun$ and a high-mass ($m_\star \approx 24.2~\msun$) giant star in a moderately eccentric ($e_\out\approx 0.46$) and tight ($P_\out\approx 14.6$ d) orbit \citep{2022Mahy}. The age of the visible star is likely 5--10 Myr, based on its mass and spectral type (O6.5\uppercase\expandafter{\romannumeral3}). Assuming an inner BHB with total mass $m_{12}=48~\msun$ and mass ratio $q=0.8$, we find that the BHB fails to excite the outer eccentricity within an integration time of 10 Myr, due to the overly compact orbit (see the left panel in Figure~\ref{fig:6}).

\item \textbf{Gaia BH1} contains a dark object of mass $\approx 9.6~\msun$ and a solar-mass ($m_\star \approx 0.93~\msun$) main-sequence (age $\gtrsim 4$ Gyr) star in a moderately eccentric ($e_\out\approx 0.45$) and wide ($\aout\approx1.4$ au) orbit \citep{gaiabh1_2023}. 
If assuming a mass ratio $q=0.8$ and an integration time of 13 Gyr, the inner BHB could excite the outer eccentricity up to $\sim0.69$ through apsidal precession resonance, and 42 out of 169 systems are stable and have high outer eccentricity ($\gtrsim0.45$) at the end of integration (see the middle panel in Figure~\ref{fig:6}).

\item \textbf{Gaia BH2} contains a dark object of mass $\approx8.9~\msun$ and a solar-mass ($m_\star\approx1~\msun$) red giant in an eccentric ($e_\out\approx 0.52$) and wide ($\aout\approx4.9$ au) orbit \citep{gaiabh2_2023}. The system age is estimated to be $\sim5$ Gyr from asteroseismology \citep{hey2025}. Adopting these parameters and again assuming a near equal-mass ($q=0.8$) inner binary, we find that the hypothesized inner BHB can excite the stellar eccentricity up to $0.68$ within 5 Gyr. All the 23 out of 247 systems that have final $\eout\gtrsim0.52$ are also stable (see the right panel in Figure~\ref{fig:6}).

\end{itemize}

In general, the apsidal precession resonance mechanism requires the inner BHB to be on a tight and eccentric orbit, {\revise which could in principal be produced by large BH natal kicks \citep[$\gtrsim100\ \text{km}\ \text{s}^{-1}$; e.g. ][]{2019MNRAS.489.3116A,2025PASP..137c4203N}.} Such a configuration can be tested by astrometric and/or RV observations, as described in Section~\ref{sec:detectability}.

\section{Conclusion} \label{sec:conclusion}
We have explored the possibility that the high eccentricity ($\approx0.73$) of the stellar orbit in Gaia BH3 system could be excited by an eccentric BHB hidden in the $33\msun$ dark component via the apsidal precession resonance mechanism. When the apsidal precession rate of the inner BHB matches that of the outer star during the orbital decay of the BHB, the system enters resonance, and the orbital eccentricity of the outer star can be excited to some extreme value. 
We show that a near equal-mass ($q=0.8$) BHB with an initial semi-major axis of 1--3 au and an initial eccentricity $\gtrsim 0.95$ is able to excite an initially nearly circular stellar orbit into the observed value ($\gtrsim0.73$) within the age of the system ($\sim$ 13 Gyr). The final semi-major axis and eccentricity of the inner BHBs that can reproduce the observed properties are {\revise $\ain \sim0.25$--0.5 au and $\ein \sim0.75$--0.85}, respectively. This mechanism can still work even if the inner and the outer orbits are moderately misaligned ($\imut\lesssim 50^\circ$).

The eccentric BHB inside Gaia BH3 is expected to induce short-term RV modulations to the visible star with semi-amplitudes of  $K_\text{short}\sim$ 10--50 ${\rm m\,s}^{-1}$ and long-term apsidal precession with rates of $\dot{\omega}_\out\sim 0.01$--$0.08$ ${\rm deg\, yr}^{-1}$. Such signals remain undetected in the currently available astrometry and RV data, but are within the reach of the full Gaia mission as well as dedicated RV follow-ups.

\begin{acknowledgments}
{\revise We thank the Editor and the anonymous referee for comments that helped to improve this paper.} Bin Liu acknowledges support from the National Key Research and Development Program of China (No. 2023YFB3002502) and National Natural Science Foundation of China (Grant No. 12433008).
Work by Q.H.\ and W.Z.\ was supported by the National Natural Science Foundation of China (grant Nos.\ 12173021 and 12133005). 
\end{acknowledgments}

\software{\texttt{numpy} \citep{harris2020array},
          \texttt{matplotlib} \citep{hunter2007matplotlib},
          \texttt{Mathematica \citep{Mathematica}}
          }


\appendix

\section{Results for Gaia BH3 System with Other Mass Ratios}\label{appendix:1}

\begin{figure*}
    \centering
    \includegraphics[width=\linewidth]{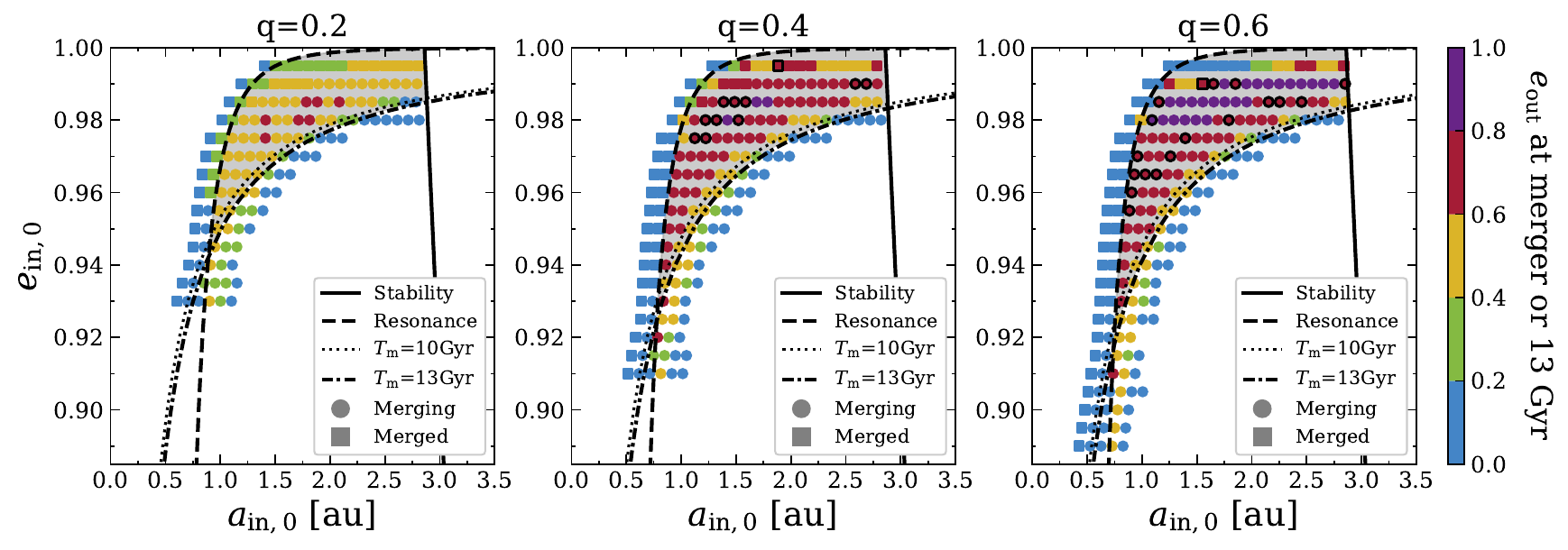}
    \caption{Similar to Figure~\ref{fig:1} but for three different mass ratios ($q=$0.2, 0.4, and 0.6).}
    \label{fig:5}
\end{figure*}

In the main text, we employ a single mass ratio of $q=0.8$ for the hypothesized BHB in the Gaia BH3 system, as this value allows the apsidal precession resonance to effectively explain the observations. However, the mass ratio of the inner BHB is a key parameter in the three conditions outlined in Section~\ref{sect:2} and significantly influences the secular evolution of the system. To ensure the completeness of this work, we supplement our findings with the results for other mass ratios.

We repeat the procedure described in Section~\ref{sect:2} for three additional mass ratios ($q=0.2,0.4,0.6$) and show the results in Figure~\ref{fig:5}, which is analogous to Figure~\ref{fig:1}. For an extreme-mass-ratio BHB ($q=0.2$), it is very difficult to excite the outer eccentricity to the observed value. As the mass ratio increases, the number of stable systems that can reproduce the high observed eccentricity ($\gtrsim0.73$) also increases, with 11 out of 186 for $q=0.4$ and 18 out of 214 for $q=0.6$ (marked out by black open circles/squares in Figure~\ref{fig:5}). This trend, however, does not extend to the case of an equal-mass BHB ($q=1.0$), where the apsidal precession resonance ceases to operate because the octupole-level interaction potential between the inner and outer orbits vanishes, as indicated by Equation (5) in \citet{liu2024extreme}.


\bibliography{ref}{}

@article{PhysRevD.96.023017,
  title = {Orbital flips in hierarchical triple systems: Relativistic effects and third-body effects to hexadecapole order},
  author = {Will, Clifford M.},
  journal = {Phys. Rev. D},
  volume = {96},
  issue = {2},
  pages = {023017},
  numpages = {15},
  year = {2017},
  month = {Jul},
  publisher = {American Physical Society},
  doi = {10.1103/PhysRevD.96.023017},
  url = {https://link.aps.org/doi/10.1103/PhysRevD.96.023017}
}

@ARTICLE{2019MNRAS.489.3116A,
       author = {{Atri}, P. and {Miller-Jones}, J.~C.~A. and {Bahramian}, A. and {Plotkin}, R.~M. and {Jonker}, P.~G. and {Nelemans}, G. and {Maccarone}, T.~J. and {Sivakoff}, G.~R. and {Deller}, A.~T. and {Chaty}, S. and {Torres}, M.~A.~P. and {Horiuchi}, S. and {McCallum}, J. and {Natusch}, T. and {Phillips}, C.~J. and {Stevens}, J. and {Weston}, S.},
        title = "{Potential kick velocity distribution of black hole X-ray binaries and implications for natal kicks}",
      journal = {\mnras},
         year = 2019,
        month = nov,
       volume = {489},
       number = {3},
        pages = {3116-3134},
          doi = {10.1093/mnras/stz2335},
archivePrefix = {arXiv},
       eprint = {1908.07199},
 primaryClass = {astro-ph.HE},
       adsurl = {https://ui.adsabs.harvard.edu/abs/2019MNRAS.489.3116A}
}

@ARTICLE{2025PASP..137c4203N,
       author = {{Nagarajan}, Pranav and {El-Badry}, Kareem},
        title = "{Mixed Origins: Strong Natal Kicks for Some Black Holes and None for Others}",
      journal = {\pasp},
         year = 2025,
        month = mar,
       volume = {137},
       number = {3},
          eid = {034203},
        pages = {034203},
          doi = {10.1088/1538-3873/adb6d6},
archivePrefix = {arXiv},
       eprint = {2411.16847},
 primaryClass = {astro-ph.GA},
       adsurl = {https://ui.adsabs.harvard.edu/abs/2025PASP..137c4203N}
}

@ARTICLE{2018ApJ...855L..15K,
       author = {{Kremer}, Kyle and {Ye}, Claire S. and {Chatterjee}, Sourav and {Rodriguez}, Carl L. and {Rasio}, Frederic A.},
        title = "{How Black Holes Shape Globular Clusters: Modeling NGC 3201}",
      journal = {\apjl},
         year = 2018,
        month = mar,
       volume = {855},
       number = {2},
          eid = {L15},
        pages = {L15},
          doi = {10.3847/2041-8213/aab26c},
archivePrefix = {arXiv},
       eprint = {1802.09553},
 primaryClass = {astro-ph.HE},
       adsurl = {https://ui.adsabs.harvard.edu/abs/2018ApJ...855L..15K}
}

@ARTICLE{2024MNRAS.527.4031T,
       author = {{Tanikawa}, Ataru and {Cary}, Savannah and {Shikauchi}, Minori and {Wang}, Long and {Fujii}, Michiko S.},
        title = "{Compact binary formation in open star clusters - I. High formation efficiency of Gaia BHs and their multiplicities}",
      journal = {\mnras},
         year = 2024,
        month = jan,
       volume = {527},
       number = {2},
        pages = {4031-4039},
          doi = {10.1093/mnras/stad3294},
archivePrefix = {arXiv},
       eprint = {2303.05743},
 primaryClass = {astro-ph.GA},
       adsurl = {https://ui.adsabs.harvard.edu/abs/2024MNRAS.527.4031T}
}

@ARTICLE{2025MNRAS.538..243F,
       author = {{Fantoccoli}, Federico and {Barber}, Jordan and {Dosopoulou}, Fani and {Chattopadhyay}, Debatri and {Antonini}, Fabio},
        title = "{Properties of black hole-star binaries formed in N-body simulations of massive star clusters: implications for Gaia black holes}",
      journal = {\mnras},
         year = 2025,
        month = mar,
       volume = {538},
       number = {1},
        pages = {243-257},
          doi = {10.1093/mnras/staf303},
archivePrefix = {arXiv},
       eprint = {2410.17323},
 primaryClass = {astro-ph.GA},
       adsurl = {https://ui.adsabs.harvard.edu/abs/2025MNRAS.538..243F}
}

@ARTICLE{Naoz2025ApJ,
       author = {{Naoz}, Smadar and {Haiman}, Zolt{\'a}n and {Quataert}, Eliot and {Holzknecht}, Liz},
        title = "{Triples as Links between Binary Black Hole Mergers, Their Electromagnetic Counterparts, and Galactic Black Holes}",
      journal = {\apjl},
         year = 2025,
        month = oct,
       volume = {992},
       number = {1},
          eid = {L12},
        pages = {L12},
          doi = {10.3847/2041-8213/ae0a20},
archivePrefix = {arXiv},
       eprint = {2508.13270},
 primaryClass = {astro-ph.HE},
       adsurl = {https://ui.adsabs.harvard.edu/abs/2025ApJ...992L..12N}
}

@ARTICLE{2025OJApEl-Badry,
       author = {{El-Badry}, Kareem and {Fabry}, Matthias and {Sana}, Hugues and {Shenar}, Tomer and {Seeburger}, Rhys},
        title = "{Complex spectral variability and hints of a luminous companion in the Be star + black hole binary candidate ALS 8814}",
      journal = {The Open Journal of Astrophysics},
         year = 2025,
        month = sep,
       volume = {8},
          eid = {128},
        pages = {128},
          doi = {10.33232/001c.143907},
       adsurl = {https://ui.adsabs.harvard.edu/abs/2025OJAp....8E.128E}
}

@ARTICLE{2022Mahy,
       author = {{Mahy}, L. and {Sana}, H. and {Shenar}, T. and {Sen}, K. and {Langer}, N. and {Marchant}, P. and {Abdul-Masih}, M. and {Banyard}, G. and {Bodensteiner}, J. and {Bowman}, D.~M. and {Dsilva}, K. and {Fabry}, M. and {Hawcroft}, C. and {Janssens}, S. and {Van Reeth}, T. and {Eldridge}, C.},
        title = "{Identifying quiescent compact objects in massive Galactic single-lined spectroscopic binaries}",
      journal = {\aap},
         year = 2022,
        month = aug,
       volume = {664},
          eid = {A159},
        pages = {A159},
          doi = {10.1051/0004-6361/202243147},
archivePrefix = {arXiv},
       eprint = {2207.07752},
 primaryClass = {astro-ph.SR},
       adsurl = {https://ui.adsabs.harvard.edu/abs/2022A&A...664A.159M}
}

@article{von1910application,
  title={Sur l'application des s{\'e}ries de M. Lindstedt {\`a} l'{\'e}tude du mouvement des com{\`e}tes p{\'e}riodiques},
  author={von Zeipel, Hugo},
  journal={Astronomische Nachrichten, volume 183, Issue 22, p. 345},
  volume={183},
  pages={345},
  year={1910}
}

@ARTICLE{1962Lidov,
       author = {{Lidov}, M.~L.},
        title = "{The evolution of orbits of artificial satellites of planets under the action of gravitational perturbations of external bodies}",
      journal = {\planss},
         year = 1962,
        month = oct,
       volume = {9},
       number = {10},
        pages = {719-759},
          doi = {10.1016/0032-0633(62)90129-0},
       adsurl = {https://ui.adsabs.harvard.edu/abs/1962P&SS....9..719L}
}

@ARTICLE{1962AJKozai,
       author = {{Kozai}, Yoshihide},
        title = "{Secular perturbations of asteroids with high inclination and eccentricity}",
      journal = {\aj},
         year = 1962,
        month = nov,
       volume = {67},
        pages = {591-598},
          doi = {10.1086/108790},
       adsurl = {https://ui.adsabs.harvard.edu/abs/1962AJ.....67..591K}
}

@article{Peters1964,
  title = {Gravitational Radiation and the Motion of Two Point Masses},
  author = {Peters, P. C.},
  journal = {Phys. Rev.},
  volume = {136},
  issue = {4B},
  pages = {B1224--B1232},
  numpages = {0},
  year = {1964},
  month = {Nov},
  publisher = {American Physical Society},
  doi = {10.1103/PhysRev.136.B1224},
  url = {https://link.aps.org/doi/10.1103/PhysRev.136.B1224}
}

@ARTICLE{Vigna2021ApJ,
       author = {{Vigna-G{\'o}mez}, Alejandro and {Toonen}, Silvia and {Ramirez-Ruiz}, Enrico and {Leigh}, Nathan W.~C. and {Riley}, Jeff and {Haster}, Carl-Johan},
        title = "{Massive Stellar Triples Leading to Sequential Binary Black Hole Mergers in the Field}",
      journal = {\apjl},
         year = 2021,
        month = jan,
       volume = {907},
       number = {1},
          eid = {L19},
        pages = {L19},
          doi = {10.3847/2041-8213/abd5b7},
archivePrefix = {arXiv},
       eprint = {2010.13669},
 primaryClass = {astro-ph.HE},
       adsurl = {https://ui.adsabs.harvard.edu/abs/2021ApJ...907L..19V}
}

@ARTICLE{Marchant2016,
       author = {{Marchant}, Pablo and {Langer}, Norbert and {Podsiadlowski}, Philipp and {Tauris}, Thomas M. and {Moriya}, Takashi J.},
        title = "{A new route towards merging massive black holes}",
      journal = {\aap},
         year = 2016,
        month = apr,
       volume = {588},
          eid = {A50},
        pages = {A50},
          doi = {10.1051/0004-6361/201628133},
archivePrefix = {arXiv},
       eprint = {1601.03718},
 primaryClass = {astro-ph.SR},
       adsurl = {https://ui.adsabs.harvard.edu/abs/2016A&A...588A..50M}
}

@ARTICLE{Vinson2018MNRAS.474.4855V,
       author = {{Vinson}, Benjamin R. and {Chiang}, Eugene},
        title = "{Secular dynamics of an exterior test particle: the inverse Kozai and other eccentricity-inclination resonances}",
      journal = {\mnras},
         year = 2018,
        month = mar,
       volume = {474},
       number = {4},
        pages = {4855-4869},
          doi = {10.1093/mnras/stx3091},
archivePrefix = {arXiv},
       eprint = {1711.10495},
 primaryClass = {astro-ph.EP},
       adsurl = {https://ui.adsabs.harvard.edu/abs/2018MNRAS.474.4855V}
}

@ARTICLE{Liu2020PhRvD.102b3020L,
       author = {{Liu}, Bin and {Lai}, Dong},
        title = "{Merging compact binaries near a rotating supermassive black hole: Eccentricity excitation due to apsidal precession resonance}",
      journal = {\prd},
         year = 2020,
        month = jul,
       volume = {102},
       number = {2},
          eid = {023020},
        pages = {023020},
          doi = {10.1103/PhysRevD.102.023020},
archivePrefix = {arXiv},
       eprint = {2004.10205},
 primaryClass = {astro-ph.HE},
       adsurl = {https://ui.adsabs.harvard.edu/abs/2020PhRvD.102b3020L}
}

@ARTICLE{Liu2022PhRvD,
       author = {{Liu}, Bin and {D'Orazio}, Daniel J. and {Vigna-G{\'o}mez}, Alejandro and {Samsing}, Johan},
        title = "{Uncovering a hidden black hole binary from secular eccentricity variations of a tertiary star}",
      journal = {\prd},
         year = 2022,
        month = dec,
       volume = {106},
       number = {12},
          eid = {123010},
        pages = {123010},
          doi = {10.1103/PhysRevD.106.123010},
archivePrefix = {arXiv},
       eprint = {2207.10091},
 primaryClass = {astro-ph.HE},
       adsurl = {https://ui.adsabs.harvard.edu/abs/2022PhRvD.106l3010L}
}

@ARTICLE{Liu2015MNRAS,
       author = {{Liu}, Bin and {Mu{\~n}oz}, Diego J. and {Lai}, Dong},
        title = "{Suppression of extreme orbital evolution in triple systems with short-range forces}",
      journal = {\mnras},
         year = 2015,
        month = feb,
       volume = {447},
       number = {1},
        pages = {747-764},
          doi = {10.1093/mnras/stu2396},
archivePrefix = {arXiv},
       eprint = {1409.6717},
 primaryClass = {astro-ph.EP},
       adsurl = {https://ui.adsabs.harvard.edu/abs/2015MNRAS.447..747L}
}

@misc{Mathematica,
  author = {Wolfram Research{,} Inc.},
  title = {Mathematica, {V}ersion 14.2},
  url = {https://www.wolfram.com/mathematica},
  note = {Champaign, IL, 2024},
  year = 2024
}

@ARTICLE{2025ALS8814,
       author = {{An}, Qian-Yu and {Huang}, Yang and {Gu}, Wei-Min and {Shao}, Yong and {Zhang}, Zhi-Xiang and {Yi}, Tuan and {Lailey}, B.~D. and {Sigut}, T.~A.~A. and {Akira Rocha}, Kyle and {Sun}, Meng and {Gossage}, Seth and {Gao}, Shi-Jie and {Weng}, Shan-Shan and {Wang}, Song and {Zhang}, Bowen and {Zhao}, Xinlin and {Qi}, Senyu and {Liao}, Shilong and {Ji}, Jianghui and {Wang}, Junfeng and {Wu}, Jianfeng and {Sun}, Mouyuan and {Li}, Xiang-Dong and {Liu}, Jifeng},
        title = "{A Be star-black hole binary with a wide orbit from LAMOST time-domain survey}",
      journal = {arXiv e-prints},
         year = 2025,
        month = may,
          eid = {arXiv:2505.23151},
        pages = {arXiv:2505.23151},
          doi = {10.48550/arXiv.2505.23151},
archivePrefix = {arXiv},
       eprint = {2505.23151},
 primaryClass = {astro-ph.SR},
       adsurl = {https://ui.adsabs.harvard.edu/abs/2025arXiv250523151A}
}

@ARTICLE{Antonini2017ApJ...841...77A,
       author = {{Antonini}, Fabio and {Toonen}, Silvia and {Hamers}, Adrian S.},
        title = "{Binary Black Hole Mergers from Field Triples: Properties, Rates, and the Impact of Stellar Evolution}",
      journal = {\apj},
         year = 2017,
        month = jun,
       volume = {841},
       number = {2},
          eid = {77},
        pages = {77},
          doi = {10.3847/1538-4357/aa6f5e},
archivePrefix = {arXiv},
       eprint = {1703.06614},
 primaryClass = {astro-ph.GA},
       adsurl = {https://ui.adsabs.harvard.edu/abs/2017ApJ...841...77A}
}

@ARTICLE{Naoz2017AJ....154...18N,
       author = {{Naoz}, Smadar and {Li}, Gongjie and {Zanardi}, Macarena and {de El{\'\i}a}, Gonzalo Carlos and {Di Sisto}, Romina P.},
        title = "{The Eccentric Kozai-Lidov Mechanism for Outer Test Particle}",
      journal = {\aj},
         year = 2017,
        month = jul,
       volume = {154},
       number = {1},
          eid = {18},
        pages = {18},
          doi = {10.3847/1538-3881/aa6fb0},
archivePrefix = {arXiv},
       eprint = {1701.03795},
 primaryClass = {astro-ph.EP},
       adsurl = {https://ui.adsabs.harvard.edu/abs/2017AJ....154...18N}
}

@ARTICLE{Tanikawa2025OJAp,
       author = {{Tanikawa}, Ataru and {Wang}, Long and {Fujii}, Michiko S. and {Trani}, Alessandro A. and {Hayashi}, Toshinori and {Suto}, Yasushi},
        title = "{Compact Binary Formation in Open Star Clusters III: Probability of Binary Black Holes Hidden Inside of Gaia Black Hole Binary}",
      journal = {The Open Journal of Astrophysics},
         year = 2025,
        month = jun,
       volume = {8},
          eid = {79},
        pages = {79},
          doi = {10.33232/001c.141294},
archivePrefix = {arXiv},
       eprint = {2407.03662},
 primaryClass = {astro-ph.GA},
       adsurl = {https://ui.adsabs.harvard.edu/abs/2025OJAp....8E..79T}
}

@ARTICLE{Iorio2024,
       author = {{Iorio}, Giuliano and {Torniamenti}, Stefano and {Mapelli}, Michela and {Dall'Amico}, Marco and {Trani}, Alessandro A. and {Rastello}, Sara and {Sgalletta}, Cecilia and {Rinaldi}, Stefano and {Costa}, Guglielmo and {Dahl-Lahtinen}, Bera A. and {Escobar}, Gast{\'o}n J. and {Korb}, Erika and {Vaccaro}, M. Paola and {Lacchin}, Elena and {Mestichelli}, Benedetta and {Di Carlo}, Ugo N. and {Spera}, Mario and {Arca Sedda}, Manuel},
        title = "{The boring history of Gaia BH3 from isolated binary evolution}",
      journal = {\aap},
         year = 2024,
        month = oct,
       volume = {690},
          eid = {A144},
        pages = {A144},
          doi = {10.1051/0004-6361/202450531},
archivePrefix = {arXiv},
       eprint = {2404.17568},
 primaryClass = {astro-ph.GA},
       adsurl = {https://ui.adsabs.harvard.edu/abs/2024A&A...690A.144I}
}

@ARTICLE{ElBadry2024OJAp,
       author = {{El-Badry}, Kareem},
        title = "{On the formation of a 33 solar-mass black hole in a low-metallicity binary}",
      journal = {The Open Journal of Astrophysics},
         year = 2024,
        month = may,
       volume = {7},
          eid = {38},
        pages = {38},
          doi = {10.33232/001c.117652},
archivePrefix = {arXiv},
       eprint = {2404.13047},
 primaryClass = {astro-ph.SR},
       adsurl = {https://ui.adsabs.harvard.edu/abs/2024OJAp....7E..38E}
}

@ARTICLE{daniel2024,
       author = {{Mar{\'\i}n Pina}, Daniel and {Rastello}, Sara and {Gieles}, Mark and {Kremer}, Kyle and {Fitzgerald}, Laura and {Rando Forastier}, Bruno},
        title = "{Dynamical formation of Gaia BH3 in the progenitor globular cluster of the ED-2 stream}",
      journal = {\aap},
         year = 2024,
        month = aug,
       volume = {688},
          eid = {L2},
        pages = {L2},
          doi = {10.1051/0004-6361/202450460},
archivePrefix = {arXiv},
       eprint = {2404.13036},
 primaryClass = {astro-ph.GA},
       adsurl = {https://ui.adsabs.harvard.edu/abs/2024A&A...688L...2M}
}

@ARTICLE{Tanikawa2023ApJ,
       author = {{Tanikawa}, Ataru and {Hattori}, Kohei and {Kawanaka}, Norita and {Kinugawa}, Tomoya and {Shikauchi}, Minori and {Tsuna}, Daichi},
        title = "{Search for a Black Hole Binary in Gaia DR3 Astrometric Binary Stars with Spectroscopic Data}",
      journal = {\apj},
         year = 2023,
        month = apr,
       volume = {946},
       number = {2},
          eid = {79},
        pages = {79},
          doi = {10.3847/1538-4357/acbf36},
archivePrefix = {arXiv},
       eprint = {2209.05632},
 primaryClass = {astro-ph.SR},
       adsurl = {https://ui.adsabs.harvard.edu/abs/2023ApJ...946...79T}
}

@ARTICLE{Chakrabarti2023AJ,
       author = {{Chakrabarti}, Sukanya and {Simon}, Joshua D. and {Craig}, Peter A. and {Reggiani}, Henrique and {Brandt}, Timothy D. and {Guhathakurta}, Puragra and {Dalba}, Paul A. and {Kirby}, Evan N. and {Chang}, Philip and {Hey}, Daniel R. and {Savino}, Alessandro and {Geha}, Marla and {Thompson}, Ian B.},
        title = "{A Noninteracting Galactic Black Hole Candidate in a Binary System with a Main-sequence Star}",
      journal = {\aj},
         year = 2023,
        month = jul,
       volume = {166},
       number = {1},
          eid = {6},
        pages = {6},
          doi = {10.3847/1538-3881/accf21},
archivePrefix = {arXiv},
       eprint = {2210.05003},
 primaryClass = {astro-ph.GA},
       adsurl = {https://ui.adsabs.harvard.edu/abs/2023AJ....166....6C}
}

@ARTICLE{Balbinot2024ED2stream,
       author = {{Balbinot}, E. and {Dodd}, E. and {Matsuno}, T. and {Lardo}, C. and {Helmi}, A. and {Panuzzo}, P. and {Mazeh}, T. and {Holl}, B. and {Caffau}, E. and {Jorissen}, A. and {Babusiaux}, C. and {Gavras}, P. and {Wyrzykowski}, {\L}. and {Eyer}, L. and {Leclerc}, N. and {Bombrun}, A. and {Mowlavi}, N. and {Seabroke}, G.~M. and {Cabrera-Ziri}, I. and {Callingham}, T.~M. and {Ruiz-Lara}, T. and {Starkenburg}, E.},
        title = "{The 33 M$_{{\ensuremath{\odot}}}$ black hole Gaia BH3 is part of the disrupted ED-2 star cluster}",
      journal = {\aap},
         year = 2024,
        month = jul,
       volume = {687},
          eid = {L3},
        pages = {L3},
          doi = {10.1051/0004-6361/202450425},
archivePrefix = {arXiv},
       eprint = {2404.11604},
 primaryClass = {astro-ph.GA},
       adsurl = {https://ui.adsabs.harvard.edu/abs/2024A&A...687L...3B}
}

@ARTICLE{gaiabh1_2023,
       author = {{El-Badry}, Kareem and {Rix}, Hans-Walter and {Quataert}, Eliot and {Howard}, Andrew W. and {Isaacson}, Howard and {Fuller}, Jim and {Hawkins}, Keith and {Breivik}, Katelyn and {Wong}, Kaze W.~K. and {Rodriguez}, Antonio C. and {Conroy}, Charlie and {Shahaf}, Sahar and {Mazeh}, Tsevi and {Arenou}, Fr{\'e}d{\'e}ric and {Burdge}, Kevin B. and {Bashi}, Dolev and {Faigler}, Simchon and {Weisz}, Daniel R. and {Seeburger}, Rhys and {Almada Monter}, Silvia and {Wojno}, Jennifer},
        title = "{A Sun-like star orbiting a black hole}",
      journal = {\mnras},
         year = 2023,
        month = jan,
       volume = {518},
       number = {1},
        pages = {1057-1085},
          doi = {10.1093/mnras/stac3140},
archivePrefix = {arXiv},
       eprint = {2209.06833},
 primaryClass = {astro-ph.SR},
       adsurl = {https://ui.adsabs.harvard.edu/abs/2023MNRAS.518.1057E}
}

@ARTICLE{gaiabh2_2023,
       author = {{El-Badry}, Kareem and {Rix}, Hans-Walter and {Cendes}, Yvette and {Rodriguez}, Antonio C. and {Conroy}, Charlie and {Quataert}, Eliot and {Hawkins}, Keith and {Zari}, Eleonora and {Hobson}, Melissa and {Breivik}, Katelyn and {Rau}, Arne and {Berger}, Edo and {Shahaf}, Sahar and {Seeburger}, Rhys and {Burdge}, Kevin B. and {Latham}, David W. and {Buchhave}, Lars A. and {Bieryla}, Allyson and {Bashi}, Dolev and {Mazeh}, Tsevi and {Faigler}, Simchon},
        title = "{A red giant orbiting a black hole}",
      journal = {\mnras},
         year = 2023,
        month = may,
       volume = {521},
       number = {3},
        pages = {4323-4348},
          doi = {10.1093/mnras/stad799},
archivePrefix = {arXiv},
       eprint = {2302.07880},
 primaryClass = {astro-ph.SR},
       adsurl = {https://ui.adsabs.harvard.edu/abs/2023MNRAS.521.4323E}
}

@ARTICLE{2019A&A...623A..45C,
       author = {{Csizmadia}, Sz. and {Hellard}, H. and {Smith}, A.~M.~S.},
        title = "{An estimate of the k$_{2}$ Love number of WASP-18Ab from its radial velocity measurements}",
      journal = {\aap},
         year = 2019,
        month = mar,
       volume = {623},
          eid = {A45},
        pages = {A45},
          doi = {10.1051/0004-6361/201834376},
archivePrefix = {arXiv},
       eprint = {1812.04463},
 primaryClass = {astro-ph.EP},
       adsurl = {https://ui.adsabs.harvard.edu/abs/2019A&A...623A..45C}
}

@ARTICLE{morais2011AAA,
       author = {{Morais}, M.~H.~M. and {Correia}, A.~C.~M.},
        title = "{Stellar wobble caused by a nearby binary system: eccentric and inclined orbits}",
      journal = {\aap},
         year = 2011,
        month = jan,
       volume = {525},
          eid = {A152},
        pages = {A152},
          doi = {10.1051/0004-6361/201014812},
archivePrefix = {arXiv},
       eprint = {1012.2375},
 primaryClass = {astro-ph.EP},
       adsurl = {https://ui.adsabs.harvard.edu/abs/2011A&A...525A.152M}
}

@ARTICLE{nagarajan2024espresso,
       author = {{Nagarajan}, Pranav and {El-Badry}, Kareem and {Triaud}, Amaury H.~M.~J. and {Baycroft}, Thomas A. and {Latham}, David and {Bieryla}, Allyson and {Buchhave}, Lars A. and {Rix}, Hans-Walter and {Quataert}, Eliot and {Howard}, Andrew and {Isaacson}, Howard and {Hobson}, Melissa J.},
        title = "{ESPRESSO Observations of Gaia BH1: High-precision Orbital Constraints and no Evidence for an Inner Binary}",
      journal = {\pasp},
         year = 2024,
        month = jan,
       volume = {136},
       number = {1},
          eid = {014202},
        pages = {014202},
          doi = {10.1088/1538-3873/ad1ba7},
archivePrefix = {arXiv},
       eprint = {2312.05313},
 primaryClass = {astro-ph.SR},
       adsurl = {https://ui.adsabs.harvard.edu/abs/2024PASP..136a4202N}
}

@ARTICLE{morias2008wobble,
       author = {{Morais}, M.~H.~M. and {Correia}, A.~C.~M.},
        title = "{Stellar wobble caused by a binary system: Can it really be mistaken as an extra-solar planet?}",
      journal = {\aap},
         year = 2008,
        month = dec,
       volume = {491},
       number = {3},
        pages = {899-906},
          doi = {10.1051/0004-6361:200810741},
archivePrefix = {arXiv},
       eprint = {0810.0506},
 primaryClass = {astro-ph},
       adsurl = {https://ui.adsabs.harvard.edu/abs/2008A&A...491..899M}
}

@ARTICLE{hayashi2020strategy,
       author = {{Hayashi}, Toshinori and {Wang}, Shijie and {Suto}, Yasushi},
        title = "{A Strategy to Search for an Inner Binary Black Hole from the Motion of the Tertiary Star}",
      journal = {\apj},
         year = 2020,
        month = feb,
       volume = {890},
       number = {2},
          eid = {112},
        pages = {112},
          doi = {10.3847/1538-4357/ab6de6},
archivePrefix = {arXiv},
       eprint = {1905.07100},
 primaryClass = {astro-ph.EP},
       adsurl = {https://ui.adsabs.harvard.edu/abs/2020ApJ...890..112H}
}

@ARTICLE{hayashi2020radial,
       author = {{Hayashi}, Toshinori and {Suto}, Yasushi},
        title = "{Radial-velocity Variation of a Tertiary Star Orbiting a Binary Black Hole in Coplanar and Noncoplanar Triples: Short- and Long-term Anomalous Behavior}",
      journal = {\apj},
         year = 2020,
        month = jul,
       volume = {897},
       number = {1},
          eid = {29},
        pages = {29},
          doi = {10.3847/1538-4357/ab97ad},
archivePrefix = {arXiv},
       eprint = {2006.02210},
 primaryClass = {astro-ph.HE},
       adsurl = {https://ui.adsabs.harvard.edu/abs/2020ApJ...897...29H}
}

@article{hunter2007matplotlib,
       author = {{Hunter}, John D.},
        title = "{Matplotlib: A 2D Graphics Environment}",
      journal = {Computing in Science and Engineering},
         year = 2007,
        month = may,
       volume = {9},
       number = {3},
        pages = {90-95},
          doi = {10.1109/MCSE.2007.55},
       adsurl = {https://ui.adsabs.harvard.edu/abs/2007CSE.....9...90H}
}

@article{harris2020array,
       author = {{Harris}, Charles R. and {Millman}, K. Jarrod and {van der Walt}, St{\'e}fan J. and {Gommers}, Ralf and {Virtanen}, Pauli and {Cournapeau}, David and {Wieser}, Eric and {Taylor}, Julian and {Berg}, Sebastian and {Smith}, Nathaniel J. and {Kern}, Robert and {Picus}, Matti and {Hoyer}, Stephan and {van Kerkwijk}, Marten H. and {Brett}, Matthew and {Haldane}, Allan and {del R{\'\i}o}, Jaime Fern{\'a}ndez and {Wiebe}, Mark and {Peterson}, Pearu and {G{\'e}rard-Marchant}, Pierre and {Sheppard}, Kevin and {Reddy}, Tyler and {Weckesser}, Warren and {Abbasi}, Hameer and {Gohlke}, Christoph and {Oliphant}, Travis E.},
        title = "{Array programming with NumPy}",
      journal = {\nat},
         year = 2020,
        month = sep,
       volume = {585},
       number = {7825},
        pages = {357-362},
          doi = {10.1038/s41586-020-2649-2},
archivePrefix = {arXiv},
       eprint = {2006.10256},
 primaryClass = {cs.MS},
       adsurl = {https://ui.adsabs.harvard.edu/abs/2020Natur.585..357H}
}

@ARTICLE{gaia2024,
       author = {{Gaia Collaboration} and {Panuzzo}, P. and {Mazeh}, T. and {Arenou}, F. and {Holl}, B. and {Caffau}, E. and {Jorissen}, A. and {Babusiaux}, C. and {Gavras}, P. and {Sahlmann}, J. and {Bastian}, U. and {Wyrzykowski}, {\L}. and {Eyer}, L. and {Leclerc}, N. and {Bauchet}, N. and {Bombrun}, A. and {Mowlavi}, N. and {Seabroke}, G.~M. and {Teyssier}, D. and {Balbinot}, E. and {Helmi}, A. and {Brown}, A.~G.~A. and {Vallenari}, A. and {Prusti}, T. and {de Bruijne}, J.~H.~J. and {Barbier}, A. and {Biermann}, M. and {Creevey}, O.~L. and {Ducourant}, C. and {Evans}, D.~W. and {Guerra}, R. and {Hutton}, A. and {Jordi}, C. and {Klioner}, S.~A. and {Lammers}, U. and {Lindegren}, L. and {Luri}, X. and {Mignard}, F. and {Nicolas}, C. and {Randich}, S. and {Sartoretti}, P. and {Smiljanic}, R. and {Tanga}, P. and {Walton}, N.~A. and {Aerts}, C. and {Bailer-Jones}, C.~A.~L. and {Cropper}, M. and {Drimmel}, R. and {Jansen}, F. and {Katz}, D. and {Lattanzi}, M.~G. and {Soubiran}, C. and {Th{\'e}venin}, F. and {van Leeuwen}, F. and {Andrae}, R. and {Audard}, M. and {Bakker}, J. and {Blomme}, R. and {Casta{\~n}eda}, J. and {De Angeli}, F. and {Fabricius}, C. and {Fouesneau}, M. and {Fr{\'e}mat}, Y. and {Galluccio}, L. and {Guerrier}, A. and {Heiter}, U. and {Masana}, E. and {Messineo}, R. and {Nienartowicz}, K. and {Pailler}, F. and {Riclet}, F. and {Roux}, W. and {Sordo}, R. and {Gracia-Abril}, G. and {Portell}, J. and {Altmann}, M. and {Benson}, K. and {Berthier}, J. and {Burgess}, P.~W. and {Busonero}, D. and {Busso}, G. and {Cacciari}, C. and {C{\'a}novas}, H. and {Carrasco}, J.~M. and {Carry}, B. and {Cellino}, A. and {Cheek}, N. and {Clementini}, G. and {Damerdji}, Y. and {Davidson}, M. and {de Teodoro}, P. and {Delchambre}, L. and {Dell'Oro}, A. and {Fraile Garcia}, E. and {Garabato}, D. and {Garc{\'\i}a-Lario}, P. and {Haigron}, R. and {Hambly}, N.~C. and {Harrison}, D.~L. and {Hatzidimitriou}, D. and {Hern{\'a}ndez}, J. and {Hestroffer}, D. and {Hodgkin}, S.~T. and {Jamal}, S. and {Jevardat de Fombelle}, G. and {Jordan}, S. and {Krone-Martins}, A. and {Lanzafame}, A.~C. and {L{\"o}ffler}, W. and {Lorca}, A. and {Marchal}, O. and {Marrese}, P.~M. and {Moitinho}, A. and {Muinonen}, K. and {Nu{\~n}ez Campos}, M. and {Oreshina-Slezak}, I. and {Osborne}, P. and {Pancino}, E. and {Pauwels}, T. and {Recio-Blanco}, A. and {Riello}, M. and {Rimoldini}, L. and {Robin}, A.~C. and {Roegiers}, T. and {Sarro}, L.~M. and {Schultheis}, M. and {Smith}, M. and {Sozzetti}, A. and {Utrilla}, E. and {van Leeuwen}, M. and {Weingrill}, K. and {Abbas}, U. and {{\'A}brah{\'a}m}, P. and {Abreu Aramburu}, A. and {Ahmed}, S. and {Altavilla}, G. and {{\'A}lvarez}, M.~A. and {Anders}, F. and {Anderson}, R.~I. and {Anglada Varela}, E. and {Antoja}, T. and {Baig}, S. and {Baines}, D. and {Baker}, S.~G. and {Balaguer-N{\'u}{\~n}ez}, L. and {Balog}, Z. and {Barache}, C. and {Barros}, M. and {Barstow}, M.~A. and {Bartolom{\'e}}, S. and {Bashi}, D. and {Bassilana}, J. -L. and {Baudeau}, N. and {Becciani}, U. and {Bedin}, L.~R. and {Bellas-Velidis}, I. and {Bellazzini}, M. and {Beordo}, W. and {Bernet}, M. and {Bertolotto}, C. and {Bertone}, S. and {Bianchi}, L. and {Binnenfeld}, A. and {Blanco-Cuaresma}, S. and {Bland-Hawthorn}, J. and {Blazere}, A. and {Boch}, T. and {Bossini}, D. and {Bouquillon}, S. and {Bragaglia}, A. and {Braine}, J. and {Bratsolis}, E. and {Breedt}, E. and {Bressan}, A. and {Brouillet}, N. and {Brugaletta}, E. and {Bucciarelli}, B. and {Butkevich}, A.~G. and {Buzzi}, R. and {Camut}, A. and {Cancelliere}, R. and {Cantat-Gaudin}, T. and {Capilla Guilarte}, D. and {Carballo}, R. and {Carlucci}, T. and {Carnerero}, M.~I. and {Carretero}, J. and {Carton}, S. and {Casamiquela}, L. and {Casey}, A. and {Castellani}, M. and {Castro-Ginard}, A. and {Ceraj}, L. and {Cesare}, V. and {Charlot}, P. and {Chaudet}, C. and {Chemin}, L. and {Chiavassa}, A. and {Chornay}, N. and {Chosson}, D.},
        title = "{Discovery of a dormant 33 solar-mass black hole in pre-release Gaia astrometry}",
      journal = {\aap},
         year = 2024,
        month = jun,
       volume = {686},
          eid = {L2},
        pages = {L2},
          doi = {10.1051/0004-6361/202449763},
archivePrefix = {arXiv},
       eprint = {2404.10486},
 primaryClass = {astro-ph.GA},
       adsurl = {https://ui.adsabs.harvard.edu/abs/2024A&A...686L...2G}
}

@article{mardling2001tidal,
       author = {{Mardling}, Rosemary A. and {Aarseth}, Sverre J.},
        title = "{Tidal interactions in star cluster simulations}",
      journal = {\mnras},
         year = 2001,
        month = mar,
       volume = {321},
       number = {3},
        pages = {398-420},
          doi = {10.1046/j.1365-8711.2001.03974.x},
       adsurl = {https://ui.adsabs.harvard.edu/abs/2001MNRAS.321..398M}
}

@article{liu2024extreme,
       author = {{Liu}, Bin and {Lai}, Dong},
        title = "{Extreme Resonant Eccentricity Excitation of Stars around Merging Black-Hole Binary}",
      journal = {\prl},
         year = 2024,
        month = jun,
       volume = {132},
       number = {23},
          eid = {231403},
        pages = {231403},
          doi = {10.1103/PhysRevLett.132.231403},
archivePrefix = {arXiv},
       eprint = {2403.03250},
 primaryClass = {astro-ph.HE},
       adsurl = {https://ui.adsabs.harvard.edu/abs/2024PhRvL.132w1403L}
}

@ARTICLE{Liu2015PhRvD,
       author = {{Liu}, Bin and {Lai}, Dong and {Yuan}, Ye-Fei},
        title = "{Merging compact binaries in hierarchical triple systems: Resonant excitation of binary eccentricity}",
      journal = {\prd},
         year = 2015,
        month = dec,
       volume = {92},
       number = {12},
          eid = {124048},
        pages = {124048},
          doi = {10.1103/PhysRevD.92.124048},
archivePrefix = {arXiv},
       eprint = {1511.07365},
 primaryClass = {astro-ph.HE},
       adsurl = {https://ui.adsabs.harvard.edu/abs/2015PhRvD..92l4048L}
}

@ARTICLE{Hayashi2023ApJ,
       author = {{Hayashi}, Toshinori and {Suto}, Yasushi and {Trani}, Alessandro A.},
        title = "{Constraining the Binarity of Black Hole Candidates: A Proof-of-concept Study of Gaia BH1 and Gaia BH2}",
      journal = {\apj},
         year = 2023,
        month = nov,
       volume = {958},
       number = {1},
          eid = {26},
        pages = {26},
          doi = {10.3847/1538-4357/acf4f6},
archivePrefix = {arXiv},
       eprint = {2307.01793},
 primaryClass = {astro-ph.HE},
       adsurl = {https://ui.adsabs.harvard.edu/abs/2023ApJ...958...26H}
}

@ARTICLE{gaia2016,
       author = {{Gaia Collaboration} and {Prusti}, T. and {de Bruijne}, J.~H.~J. and {Brown}, A.~G.~A. and {Vallenari}, A. and {Babusiaux}, C. and {Bailer-Jones}, C.~A.~L. and {Bastian}, U. and {Biermann}, M. and {Evans}, D.~W. and {Eyer}, L. and {Jansen}, F. and {Jordi}, C. and {Klioner}, S.~A. and {Lammers}, U. and {Lindegren}, L. and {Luri}, X. and {Mignard}, F. and {Milligan}, D.~J. and {Panem}, C. and {Poinsignon}, V. and {Pourbaix}, D. and {Randich}, S. and {Sarri}, G. and {Sartoretti}, P. and {Siddiqui}, H.~I. and {Soubiran}, C. and {Valette}, V. and {van Leeuwen}, F. and {Walton}, N.~A. and {Aerts}, C. and {Arenou}, F. and {Cropper}, M. and {Drimmel}, R. and {H{\o}g}, E. and {Katz}, D. and {Lattanzi}, M.~G. and {O'Mullane}, W. and {Grebel}, E.~K. and {Holland}, A.~D. and {Huc}, C. and {Passot}, X. and {Bramante}, L. and {Cacciari}, C. and {Casta{\~n}eda}, J. and {Chaoul}, L. and {Cheek}, N. and {De Angeli}, F. and {Fabricius}, C. and {Guerra}, R. and {Hern{\'a}ndez}, J. and {Jean-Antoine-Piccolo}, A. and {Masana}, E. and {Messineo}, R. and {Mowlavi}, N. and {Nienartowicz}, K. and {Ord{\'o}{\~n}ez-Blanco}, D. and {Panuzzo}, P. and {Portell}, J. and {Richards}, P.~J. and {Riello}, M. and {Seabroke}, G.~M. and {Tanga}, P. and {Th{\'e}venin}, F. and {Torra}, J. and {Els}, S.~G. and {Gracia-Abril}, G. and {Comoretto}, G. and {Garcia-Reinaldos}, M. and {Lock}, T. and {Mercier}, E. and {Altmann}, M. and {Andrae}, R. and {Astraatmadja}, T.~L. and {Bellas-Velidis}, I. and {Benson}, K. and {Berthier}, J. and {Blomme}, R. and {Busso}, G. and {Carry}, B. and {Cellino}, A. and {Clementini}, G. and {Cowell}, S. and {Creevey}, O. and {Cuypers}, J. and {Davidson}, M. and {De Ridder}, J. and {de Torres}, A. and {Delchambre}, L. and {Dell'Oro}, A. and {Ducourant}, C. and {Fr{\'e}mat}, Y. and {Garc{\'\i}a-Torres}, M. and {Gosset}, E. and {Halbwachs}, J. -L. and {Hambly}, N.~C. and {Harrison}, D.~L. and {Hauser}, M. and {Hestroffer}, D. and {Hodgkin}, S.~T. and {Huckle}, H.~E. and {Hutton}, A. and {Jasniewicz}, G. and {Jordan}, S. and {Kontizas}, M. and {Korn}, A.~J. and {Lanzafame}, A.~C. and {Manteiga}, M. and {Moitinho}, A. and {Muinonen}, K. and {Osinde}, J. and {Pancino}, E. and {Pauwels}, T. and {Petit}, J. -M. and {Recio-Blanco}, A. and {Robin}, A.~C. and {Sarro}, L.~M. and {Siopis}, C. and {Smith}, M. and {Smith}, K.~W. and {Sozzetti}, A. and {Thuillot}, W. and {van Reeven}, W. and {Viala}, Y. and {Abbas}, U. and {Abreu Aramburu}, A. and {Accart}, S. and {Aguado}, J.~J. and {Allan}, P.~M. and {Allasia}, W. and {Altavilla}, G. and {{\'A}lvarez}, M.~A. and {Alves}, J. and {Anderson}, R.~I. and {Andrei}, A.~H. and {Anglada Varela}, E. and {Antiche}, E. and {Antoja}, T. and {Ant{\'o}n}, S. and {Arcay}, B. and {Atzei}, A. and {Ayache}, L. and {Bach}, N. and {Baker}, S.~G. and {Balaguer-N{\'u}{\~n}ez}, L. and {Barache}, C. and {Barata}, C. and {Barbier}, A. and {Barblan}, F. and {Baroni}, M. and {Barrado y Navascu{\'e}s}, D. and {Barros}, M. and {Barstow}, M.~A. and {Becciani}, U. and {Bellazzini}, M. and {Bellei}, G. and {Bello Garc{\'\i}a}, A. and {Belokurov}, V. and {Bendjoya}, P. and {Berihuete}, A. and {Bianchi}, L. and {Bienaym{\'e}}, O. and {Billebaud}, F. and {Blagorodnova}, N. and {Blanco-Cuaresma}, S. and {Boch}, T. and {Bombrun}, A. and {Borrachero}, R. and {Bouquillon}, S. and {Bourda}, G. and {Bouy}, H. and {Bragaglia}, A. and {Breddels}, M.~A. and {Brouillet}, N. and {Br{\"u}semeister}, T. and {Bucciarelli}, B. and {Budnik}, F. and {Burgess}, P. and {Burgon}, R. and {Burlacu}, A. and {Busonero}, D. and {Buzzi}, R. and {Caffau}, E. and {Cambras}, J. and {Campbell}, H. and {Cancelliere}, R. and {Cantat-Gaudin}, T. and {Carlucci}, T. and {Carrasco}, J.~M. and {Castellani}, M. and {Charlot}, P. and {Charnas}, J. and {Charvet}, P. and {Chassat}, F. and {Chiavassa}, A. and {Clotet}, M. and {Cocozza}, G. and {Collins}, R.~S. and {Collins}, P. and {Costigan}, G.},
        title = "{The Gaia mission}",
      journal = {\aap},
         year = 2016,
        month = nov,
       volume = {595},
          eid = {A1},
        pages = {A1},
          doi = {10.1051/0004-6361/201629272},
archivePrefix = {arXiv},
       eprint = {1609.04153},
 primaryClass = {astro-ph.IM},
       adsurl = {https://ui.adsabs.harvard.edu/abs/2016A&A...595A...1G}
}

@ARTICLE{hey2025,
       author = {{Hey}, Daniel and {Li}, Yaguang and {Ong}, Joel},
        title = "{Asteroseismology of the red giant companions to Gaia BH2 and BH3}",
      journal = {arXiv e-prints},
         year = 2025,
        month = mar,
          eid = {arXiv:2503.09690},
        pages = {arXiv:2503.09690},
          doi = {10.48550/arXiv.2503.09690},
archivePrefix = {arXiv},
       eprint = {2503.09690},
 primaryClass = {astro-ph.SR},
       adsurl = {https://ui.adsabs.harvard.edu/abs/2025arXiv250309690H}
}
\bibliographystyle{aasjournalv7}



\end{CJK*}
\end{document}